\documentclass[superscriptaddress,twocolumn,amsmath,amssymb,aps,prb,floatfix]{revtex4-1}
\usepackage{graphicx,color}
\usepackage{dcolumn}
\usepackage{bm,verbatim}
\usepackage{hyperref}
\usepackage[mathlines]{lineno}

\begin{document}
	
	\title{Out-of-time-ordered commutators in Dirac--Weyl systems}
	
	\author{Z. Okv\'{a}tovity}
	\email{okvatovity@phy.bme.hu}
	\affiliation{Department of Theoretical Physics and MTA-BME Lend\"{u}let Topology and Correlation Research Group, Budapest University of Technology and Economics,1521 Budapest, Hungary}
	
	\author{B. D\'{o}ra}
	
	\affiliation{Department of Theoretical Physics and MTA-BME Lend\"{u}let Topology and Correlation Research Group, Budapest University of Technology and Economics,1521 Budapest, Hungary}
	
	\date{\today}
	
	\begin{abstract}        
		Quantum information stored in local operators spreads over other degrees of freedom of the system during time evolution, known as scrambling. This process is conveniently characterized by the out-of-time-order 
commutators (OTOC), whose time dependence reveals salient aspects of the system's dynamics. Here we study the spatially local spin correlation function i.e., the expectation value of spin commutator and the corresponding OTOC 
of Dirac--Weyl systems in 1, 2 and 3 spatial dimensions. The OTOC can be written as the square of the expectation 
value of the commutator and the variance of the commutator. In principle, the problem features two energy scales, 
the chemical potential, and the high energy cutoff. We find that only the latter is dominant, therefore the time 
evolution is separated into only two different regions. The spin correlation function grows linearly with time 
initially and decays as $t^{-2}$  for late times. The OTOC reveals a universal $t^2$ initial growth from both the 
commutator and the variance while its late time decay, $t^{-2}$ originates from the variance of the commutator. 
This late time decay is identified as a characteristic signature or Dirac-Weyl fermions.
	These features remain present also at finite temperatures.
		Our results indicate that Dirac--Weyl systems are slow information scramblers and are essential 
when additional channels for scrambling, i.e. interaction or disorder are analyzed.
	\end{abstract}
	
	\maketitle    
	
	\section{\label{sec:introduction}Introduction}
	In recent studies, the chaotic behavior of quantum systems has been investigated from different viewpoints\cite{Steinberg2019,Roberts2016,Cotler2017}. A common property of chaotic systems is that during unitary time evolution, simple operators can become highly complicated. This leads to the scrambling of information stored in local operator\cite{Zhou2018}. This phenomena is investigated in different contexts i.e., random matrix theory\cite{Bohigas1984}, black holes\cite{Cotler2017} and quantum thermalisation\cite{Srednicki1994,Deutsch1991}.
	
	There are multiple ways to characterize chaos and information scrambling e.g. operator entanglement entropy or out-of-time-ordered commutator (OTOC)\cite{Prosen2007,Hashimoto2017}. In this paper, we  focus on the latter which was originally introduced by Larkin and Ovchinnikov in 1969 in a calculation of the non-linear correction to the conductivity of a dirty superconductor \cite{Ovchinnikov1969}. The OTOC can be considered as the second moment of the commutator, defined as     
	\begin{equation}
	C(t)=-\left\langle\left[W(t),V(0)\right]^2\right\rangle\geq 0,
	\label{eq:otocgen}
	\end{equation}
	where $W$ and $V$ are local operators possibly separated by finite distance and $W(t)=\exp(iHt)W\exp(-iHt)$ denotes the Heisenberg time evolution. 
	
	The OTOC is a useful tool to measure the sensitivity of the time evolution of the system on the initial conditions\cite{Roberts2016} and to characterize information spreading processes. The information stored in local operators are spreading over many degrees of freedom during the time evolution and cannot be restored by local measurements\cite{Page1993}. This process i.e., the loss of information through delocalization is called scrambling\cite{Lashkari2013,Sekino2008}. In systems, where the short time exponential growth, bounded by thermal Lyapunov exponent $\lambda_L\leq 2\pi k_BT$ \cite{Roberts2016,Maldacena2016,Tsuji2018} is present are called fast scramblers, but there are also models where this short time growth is absent, called slow scramblers. Thus, a possible way to understand the nature of chaos and test capabilities of condensed matter systems for quantum information processing is to investigate the temporal behavior of OTOC. It has already been analyzed in a variety of systems, including Luttinger liquids\cite{Dra2017}, random unitary circuits\cite{Nahum2018,Rakovszky2018,vonKeyserlingk2018} in quantum Ising chain\cite{Lin2018}, XY chains\cite{Jiahui2019}, conformal field theories \cite{Roberts2015,Stanford2016} and Sachdev-Ye-Kitaev model\cite{Sachdev1993,Maldacena20162,Tsuji2017}. OTOC has been investigated experimentally as well in different many-body systems such as in cold atomic systems, trapped ions or in a nuclear magnetic resonance quantum simulator\cite{Li2017,Swingle2016,Zhu2016,Kaufman2016,Grttner2017,Landsman2019,Xinfang2019}.     
	
	In this paper, we consider Dirac--Weyl systems in one, two and three dimensions, characterized by linear energy-momentum relation. These models are popular not only in the condensed matter physics but possess a rich history in high energy physics as well\cite{Kharzeev2009}. As already mentioned, the common property of these systems is the linear energy-momentum dispersion relation and the Brillouin zone contains monopole-like structures called Weyl or Dirac nodes. The most famous descendants are carbon nanotubes in one, graphene in two and Weyl-semimetals in three dimensions \cite{Armitage2018,CastroNeto2009}, respectively. The low energy excitations are described by massless non-interacting fermions and one of their unique features is that they can host topologically non-trivial states which are robust against small perturbations \cite{Okuyama2018,Hasan2010,Armitage2018}. This non-trivial topology shows up in exotic electromagnetic transport phenomena such as topologically protected edge states in zigzag carbon nanotubes in presence of spin-orbit interaction\cite{Okuyama2018}, the quantum spin Hall effect in graphene\cite{Kane2005} or the chiral anomaly \cite{Goswami2013,Vishwanath2013} or the anomalous Hall conductivity in Weyl semimetals \cite{Burkov2011}.
	
	To obtain the short and late time behavior of the OTOC, we rewrite Eq. (\ref{eq:otocgen}) in a more suggestive way as 
	
	\begin{equation}
	C(t)=-\left\langle\left[W(t),V(0)\right]\right\rangle^2+K(t)
	\label{eq:otocgen2}
	\end{equation}
	where the first term is the square of the expectation value of the spin commutator and $K(t)$ is the variance of the commutator, i.e. $K(t)=\left\langle\left[W(t),V(0)\right]\right\rangle^2-\left\langle\left[W(t),V(0)\right]^2\right\rangle$. In other words, $K(t)$ measures the spreading of the probability distribution of the commutator compared to its expectation value. When the commutator as an operator is a c-number, the variance is zero\cite{Dra2017} which means the distribution of the commutator is a Dirac-delta function. A finite variance is the first indicator of the broadening of the distribution around the mean value. In our case, this indicates how the OTOC, i.e. the expectation value of the square of the commutator differs from the square of the expectation value of the commutator. One can in principle also obtain the full counting statistics of the commutator in general by calculating higher moments as $\langle\left[W(t), V(0)\right]^n \rangle$ similarly to  Eq. (\ref{eq:otocgen}). 
	
	In this work, we investigate the short and late time behavior of OTOC in Dirac--Weyl systems. In this context, the OTOC was already investigated in interacting graphene\cite{klug} Weyl semimetals\cite{Chen2019} from different approaches. We focus on the response of a single non-interacting Dirac--Weyl cone which is diagonalizable straightforwardly, thus integrable. We focus on the correlators of the spin and density operators, and the former is also directly proportional to the electric current operator in Dirac--Weyl systems. These correlators are necessary to characterize the electric and magnetic properties of these systems \cite{Zhou20182}. The short time behavior of OTOC gives a universal $t^2$ initial rise from both the square of the expectation value of the commutator and the variance from Eq. \eqref{eq:otocgen2}, though the contribution of the former parametrically is dominant over the latter. The late time decay is dominated only by $K(t)$ and depends on the dimension of the system and is independent of the chemical potential. 
	
	This paper is organized as follows: in Sec. \ref{sec:hamiltonian},  the model Hamiltonian and the operators are introduced. In Sec. \ref{sec:respose}, the time-dependent expectation value of the spin commutator is calculated by separating the matrix element and occupation number dependent parts. In Sec. \ref{sec:otoc} calculation procedure of OTOC is outlined and briefly discuss the results. In Sec. \ref{sec:summary}, our main results are summarized.
	
	\section{\label{sec:hamiltonian}Low energy effective Hamiltonian}
	
	The low energy effective Hamiltonian of Dirac--Weyl systems in $d$ dimension is given by    
	
	\begin{equation}
	H=v_\text{F}\mathbf{p}\cdot\boldsymbol{\sigma}
	\label{eq:ham}
	\end{equation}
	where $\textbf{p}$ is the momentum operator, $v_\text{F}$ is the Fermi velocity and $\boldsymbol{\sigma}$ denotes the corresponding Pauli matrices. 
For $d=1$ only $\sigma_{x}$ appears in the Hamiltonian, for $d=2$, $\boldsymbol{\sigma}=\left[\sigma_{x},\sigma_{y}\right]$ and for three dimensions all Pauli 
matrices are present. At low energy, the energy dispersion relation is linear in momentum for any dimension: 
$\varepsilon_{\lambda}(\mathbf{k})=\lambda\hbar v_\text{F}|\mathbf{k}|$ with $\lambda=\pm$ the band index\cite{CastroNeto2009,Armitage2018}. The corresponding wavefunctions are written as     
	
	\begin{equation}
	\phi_{\lambda,\mathbf{k}}(\mathbf{r})=\frac{1}{\sqrt{V_{d}}}e^{i\mathbf{kr}}\left|k,\lambda\right\rangle,
	\label{eq:wavefunc}
	\end{equation}
	where $V_{d}$ is the volume and  $\left|k,\lambda\right\rangle$ is the normalized eigenspinor written in $d$ dimension as        
	
	\begin{gather}
		\left|k,\lambda\right\rangle=
		\begin{cases}
			\frac{1}{\sqrt{2}}
			\begin{bmatrix}
				1 \\
				\lambda\text{sgn}(k)
			\end{bmatrix} &\text{for } d=1,\vspace{1mm}\\
			\frac{1}{\sqrt{2}}
			\begin{bmatrix}
				\lambda \\
				e^{i\varphi_{k}}
			\end{bmatrix} &\text{for } d=2,\vspace{1mm}\\
			\begin{bmatrix}
				\cos{\left(\frac{\vartheta_{k}+(\lambda-1)\pi/2}{2}\right)} \\
				\sin{\left(\frac{\vartheta_{k}+(\lambda-1)\pi/2}{2}\right)}e^{i\varphi_{ k}} \\    
			\end{bmatrix} &\text{for } d=3.    
		\end{cases}
		\label{eq:spinors}    
	\end{gather}
	Here $\varphi_{k}$ is the polar angle in two and the azimuthal angle in three dimensions and $\vartheta_{k}$ is the polar angle in 3D. Using Eq. (\ref{eq:wavefunc}), we construct the local field operators. Considering a Hermitian operator $\mathcal{O}$, the spatial and time dependent formula in second quantized formalism is given by\cite{Mahan2000} 
	
	\begin{equation}
	\mathcal{O}({\bf r},t)=\frac{1}{V_{d}}\sum_{k_{1},k_{2}}e^{i{\bf (k_1-k_2)r}}T(k_1,k_2)
	\left\langle k_2|\mathcal{O}|k_1\right\rangle c^{\dagger}_{k_2}c_{k_1}.
	\label{eq:secop}
	\end{equation}
	Here,  $k_{i}=({\bf k_{i}},\lambda_i)$ is a combination of the momentum state and band index, and  $c^{\dagger}_{k_2}$ and $c_{k_1}$ are fermionic creation and annihilation operators into state $k_2$ and $k_1$, respectively. The time dependence is represented by $T(k_1,k_2)=\exp\left[i(\varepsilon_{\lambda_2}(\mathbf{k}_2)-\varepsilon_{\lambda_1}(\mathbf{k}_1))t\right]$ with $\varepsilon_{\lambda_i}(\mathbf{k}_i)$, the energy of the Dirac--Weyl fermions and $\left\langle k_2|\mathcal{O}|k_1\right\rangle$ is the corresponding matrix element of $\mathcal{O}$ operator. In the following sections, we  focus on the time dependence of spin and density commutators and the corresponding OTOC, thus we set $\textbf{r}=0$ in Eq. (\ref{eq:secop}) during the calculations. 
	We find that for short times the commutator and the OTOC grow with $t$ and $t^2$, respectively. For late times, both quantities decay as $t^{-2}$. To corroborate these results, we also consider the spatial dependence in one dimension analytically. We find that the initial growth and late time decay remain intact, and based on these results, we expect  similar temporal behavior in higher dimensions.
	
	\section{\label{sec:respose} Correlation function}
	The correlation function of spin or density operators $\sigma_{\alpha}(\textbf{r},t)$ stores important information about the electric and magnetic properties of Dirac--Weyl systems since the former is directly proportional to the electric current operator. Within the framework of linear response it is  defined by the standard Kubo formula as\cite{Giuliani2005} 
	
	\begin{equation}
		\Pi^{\alpha\beta}(\textbf{r},t)=i\left\langle\left[\sigma_{\alpha}(\textbf{r},t),\sigma_{\beta} (0,0)\right]\right\rangle\Theta(t).
		\label{eq:kubo}
		\end{equation}
	Here, $\Theta(t)$ is the Heaviside step function and $\alpha,\beta=0,x,y,z$. For diagonal case, when $\alpha=\beta$ the expectation value of the commutator is
	
	\begin{eqnarray}
		\Pi^{\alpha\alpha}(\textbf{r},t)= \frac{i}{V_{d}^2}\sum_{k_1,k_2}e^{i{\bf (k_1-k_2)r}}T(k_1,k_2)\times\nonumber\\
		\times\left|\left\langle k_2|\sigma_{\alpha}|k_1\right\rangle\right|^2\left[f(k_2)-f(k_1)\right],   
		\label{eq:kuboexp}
		\end{eqnarray}
	for $t>0$ and $f(k_i)=\left[\exp((\varepsilon_{\lambda_i}(\mathbf{k}_i)-\mu)/k_\text{B} T)+1\right]^{-1}$ is the Fermi function. First, we focus on the time dependence of the correlation function, so we set $\textbf{r}=0$. In this case, Eq. (\ref{eq:kuboexp}) consists of two terms:  the occupation number dependent part and the absolute value square of the matrix elements which are treated separately in the followings.
	
	\subsection{\label{sec:matrix}Matrix elements}        
	By rewriting the appropriate polar and azimuthal coordinates of the eigenspinors in Eq. (\ref{eq:spinors}) to Cartesian coordinates of momentum vector, the square of matrix elements in Eq. (\ref{eq:kuboexp}) has a closed formula in $d$ dimension given by 
	
	\begin{align}
			\left|\left\langle k_2|\sigma_{0}|k_1\right\rangle\right|^2&=\frac{1}{2}\left[1+\lambda_1 \lambda_2\frac{{\bf k_1 k_2}}{|{\bf k_1}||{\bf k_2}|}\right]
			\label{eq:matsquare1}\\
			\left|\left\langle k_2|\sigma_{i}|k_1\right\rangle\right|^2&=\frac{1}{2}\left[1-\lambda_1 \lambda_2\frac{{\bf k_1 k_2}-2k_{1,i}k_{2,i}}{|{\bf k_1}||{\bf k_2}|}\right].
			\label{eq:matsquare2}
	\end{align}
	Rewriting the summations into spherical integrals in Eq. (\ref{eq:kuboexp}), we  evaluate the angular integrals separately since only the square of matrix elements in Eq. \eqref{eq:matsquare1} and \eqref{eq:matsquare2} depend on angular variables and are independent from the absolute value of momentum. In two and three dimensions, the integrals over angular variables are  
	
	\begin{equation}
	M^{\alpha}_d=\int\frac{d\Omega_1 d\Omega_2}{(2\pi)^{2(d-1)}}\left|\left\langle k_2|\sigma_{\alpha}|k_1\right\rangle\right|^2,
	\label{eq:intmatsquare}
	\end{equation}
	where $d\Omega_i=d\varphi_{k_i}$ in two and $d\Omega_i=d\varphi_{k_i}\sin(\vartheta_{k_i})d\vartheta_{k_i}$ in three dimensions with $i=1,2$. In one dimension, there is no angular integral, but we can still introduce $M^{\alpha}_1$ by changing the boundary of the momentum integrals from $(-\infty,\infty)$ to $[0,\infty)$. Evaluating the integrals, we end up with    
	
	\begin{equation}
	M^{\alpha}_1=2, ~M^{\alpha}_2=\frac{1}{2}, ~M^{\alpha}_3=\frac{2}{(2\pi)^2}.
	\label{eq:mat}
	\end{equation}
	The angular integral of the square of matrix elements yields a band index ($\lambda$) independent constant that we use in the next section to derive the time dependence of the expectation value of the commutator.
	
	\subsection{\label{sec:expresp}Explicit form of the expectation value of the commutator}    
	We evaluate the remaining integrals over the radial component of momentum. Let us introduce the following functions as   
	
	\begin{align}
			G_{d}^{<}(t)&=\sum_{\lambda}\int_{0}^{\Lambda}\frac{dk}{2\pi}k^{d-1}e^{-i\varepsilon_{\lambda}(k)t}f(k)
			\label{eq:greenless}\\
			G_{d}^{>}(t)&=\sum_{\lambda}\int_{0}^{\Lambda}\frac{dk}{2\pi}k^{d-1}e^{-i\varepsilon_{\lambda}(k)t}(1-f(k)).
			\label{eq:greengreat}
	\end{align}
	Here, $dk$ denotes the integral over the radial component of the momentum and  $\Lambda$ is sharp cutoff in momentum which leads to a sharp energy cutoff as $W=v_\text{F}\Lambda$. This type of regularization is typical in condensed matter physics and arise in e.g. Brillouin zone integrals and tight-binding models \cite{Lin20182}. The functions defined in Eqs. (\ref{eq:greenless}) and (\ref{eq:greengreat}) are practically the greater and lesser Green’s functions\cite{Mahan2000, Rammer1986} with the only difference that instead of summing over the full momentum space, only the radial component is integrated over.     
	We express the expectation value of the commutator in terms of these Green's functions as 
	
	\begin{equation}
	\Pi^{\alpha\alpha}(t)=-2M^{\alpha}_d\text{Im}\left[G_{d}^{>}(t)\bar{G}_{d}^{<}(t)\right].
	\label{eq:kubogreen}
	\end{equation}
	We are interested in the zero temperature limit, allowing us to substitute the Fermi function with a Heaviside step function. In these non-interacting systems, no exponential growth with thermal Lyapunov exponent is expected, which justifies this simplification, though the temperature dependence will be commented on later. The integral  in the Green's functions is evaluated in Appendix \ref{app:appa}. After some straightforward algebra, the commutator looks as    
	
	\begin{widetext}    
			\begin{gather}
				\Pi^{\alpha\alpha}(\tau)=
				\begin{cases}
					\dfrac{8\Lambda^2}{(2\pi)^2}\dfrac{1}{\tau^2}\sin(\tau)\left[\cos(\nu \tau)- \cos(\tau)\right] &\text{for } d=1,\vspace{2mm}\\
					\dfrac{2\Lambda^4}{(2\pi)^2}\dfrac{1}{\tau^4}\left(\cos(\tau)+\tau \sin(\tau)-1\right)\left(\sin(\tau)-\tau\cos(\tau)-\left(\sin(\nu\tau)-\nu\tau\cos(\nu \tau)\right)\right)
					&\text{for } d=2,\vspace{2mm}\\
					\dfrac{8\Lambda^6}{ (2\pi)^4}\dfrac{1}{\tau^6}\left(2\tau\cos(\tau)-(2-\tau^2)\sin(\tau)
					\right)\Big((2-\tau^2)\cos(\tau)+2\tau\sin(\tau)-\\-\left((2-\nu^2\tau^2)\cos(\nu\tau)+2\tau\sin(\nu\tau)\right)
					\Big) &\text{for } d=3.    
				\end{cases}
				\label{eq:kuboe}    
			\end{gather}        
	\end{widetext}
	Here, dimensionless variables for time as $\tau= \Lambda v_\text{F}t$ and  $\nu=|\mu|/\Lambda v_\text{F}$ for chemical potential are introduced. 
	
\begin{figure}[!ht]
		\includegraphics[width=\linewidth]{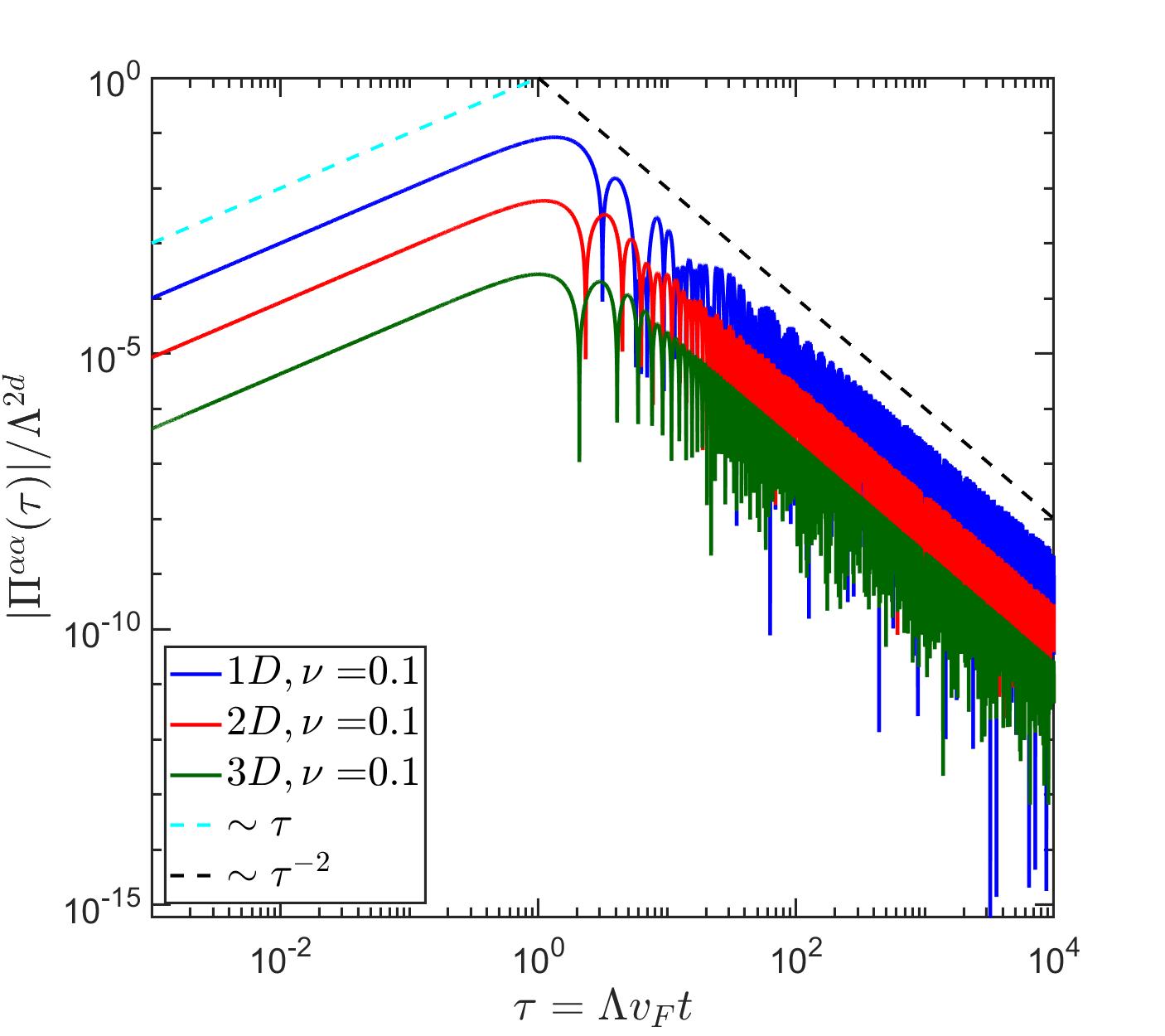}
		\caption{\label{fig:kuboshort} The temporal dynamics of the spin and density commutator in one (blue), two (red) and three (green) dimensions are plotted. In any dimension, the early time growth scales with $\tau$ (light blue dashed line). For late times, the response decays as a power-law as $\tau^{-2}$ (black dashed line). Both the early and late time growth are independent of the chemical potential and spatial dimension.}
	\end{figure}

	The spin correlation functions feature two energy scales the chemical potential and the high energy cutoff.  Thus, we expect three distinct regions with respect to time: short time growth for $\tau \ll 1$, intermediate times when $1\ll\tau \ll 1/\nu$ and the late time decay when $\tau \gg 1$. We find that only the cutoff related timescale matters and it is enough to focus on two temporal regions. Since the chemical potential always lies below the high energy cutoff, we use  $\nu\ll 1$. For short times, the expectation value of the commutator grows linearly with time in all considered dimensions, depicted in Fig. \ref{fig:kuboshort} as 
	
	\begin{equation}
		\Pi^{\alpha\alpha}(t\to 0)\sim \Lambda^dt\left((\Lambda v_\text{F})^{d+1}-\mu^{d+1}\right).
		\label{eq:kuboshort}
		\end{equation} 
		The slope of the linear growth is determined by expectation value of the energy of the states between the chemical potential and the cutoff which is coming from the first order expansion of the time evolution operator. Due to $\Lambda v_\text{F}\gg\mu$, the cutoff dependence is dominant.     
	
	The other case, when $\tau \gg1$, the asymptotic behavior is a $t^{-2}$ power-law decay in all dimension given by  
	
	\begin{eqnarray}
		\Pi^{\alpha\alpha}(t\to \infty)\sim \frac{\Lambda^{d-1}\sin(\Lambda v_\text{F}t)}{t^2}\times \nonumber\\
		\times \left[\mu^{d-1}\cos(\mu t)-(\Lambda v_\text{F})^{d-1}\cos(\Lambda v_\text{F}t)\right].
		\label{eq:kubolate}
		\end{eqnarray}
		The prefactor of the decay is determined by the density of states, $\rho(\varepsilon)$ since in  Dirac--Weyl systems $\rho(\varepsilon)\sim \varepsilon^{d-1}$. The late time behavior is also dominated  by the cutoff and the same reasoning holds as in the short time case. We displayed the short and and late time behaviors of the expectation value of the spin and density commutator in Fig. \ref{fig:kuboshort}.    
	
	We argue that the $t^{-2}$ decay for late times, independent from the dimension $d$, is a characteristic signature of Dirac-Weyl fermions. Such systems are
characterized by linear energy-momentum relation, i.e. $\varepsilon_{\lambda}(\mathbf{k})\sim |\mathbf{k}|$. By retaining this relationship in the time dependent factor 
$T(k_1,k_2)$ in 
Eq. \eqref{eq:kuboexp} and neglecting the wavevector dependence of all other quantities, we end up with the very same decay as in Eq. \eqref{eq:kubolate} after
performing the $d$-dimensional momentum integral.

\begin{figure}[!ht]
		\includegraphics[width=\linewidth]{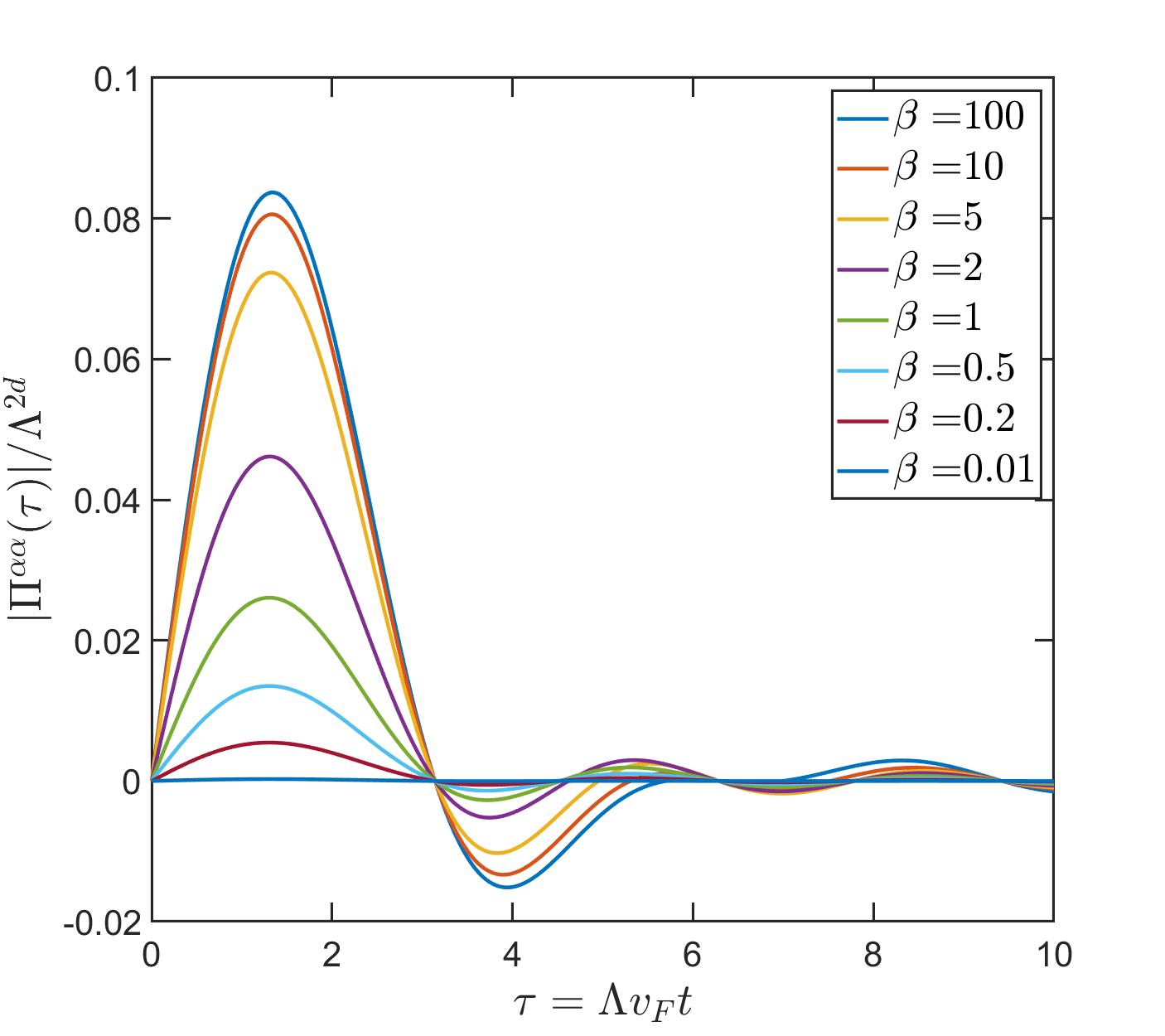}
		\caption{\label{fig:kubotemp} The temporal dynamics of the spin and density commutator is plotted on different temperatures in one  dimension. $\beta=\Lambda v_\text{F}/k_{B}T$ denotes the inverse temperature. Increasing the temperature ($\beta\to 0$), the main characteristics of $\Pi^{\alpha\alpha}(\tau)$ remains the same, but the amplitude decrease. At the $\beta=0$ limit the commutator vanishes. We set $\nu=0.1$ for the numerical calculation.}
	\end{figure}

	We also compute numerically the temperature dependence of the expectation value of the commutator by keeping the general form of Fermi function. The result is plotted in Fig. \ref{fig:kubotemp}.  The numerical calculation showed the magnitude of the commutator decrease as a smooth function of temperature, but the short and late time behavior remains the same. In the $T\to \infty$ limit, the Green's functions in Eqs. \eqref{eq:greenless} and \eqref{eq:greengreat} are real, so the  expectation value of the commutator vanishes.

\begin{figure}[!ht]
		\includegraphics[width=\linewidth]{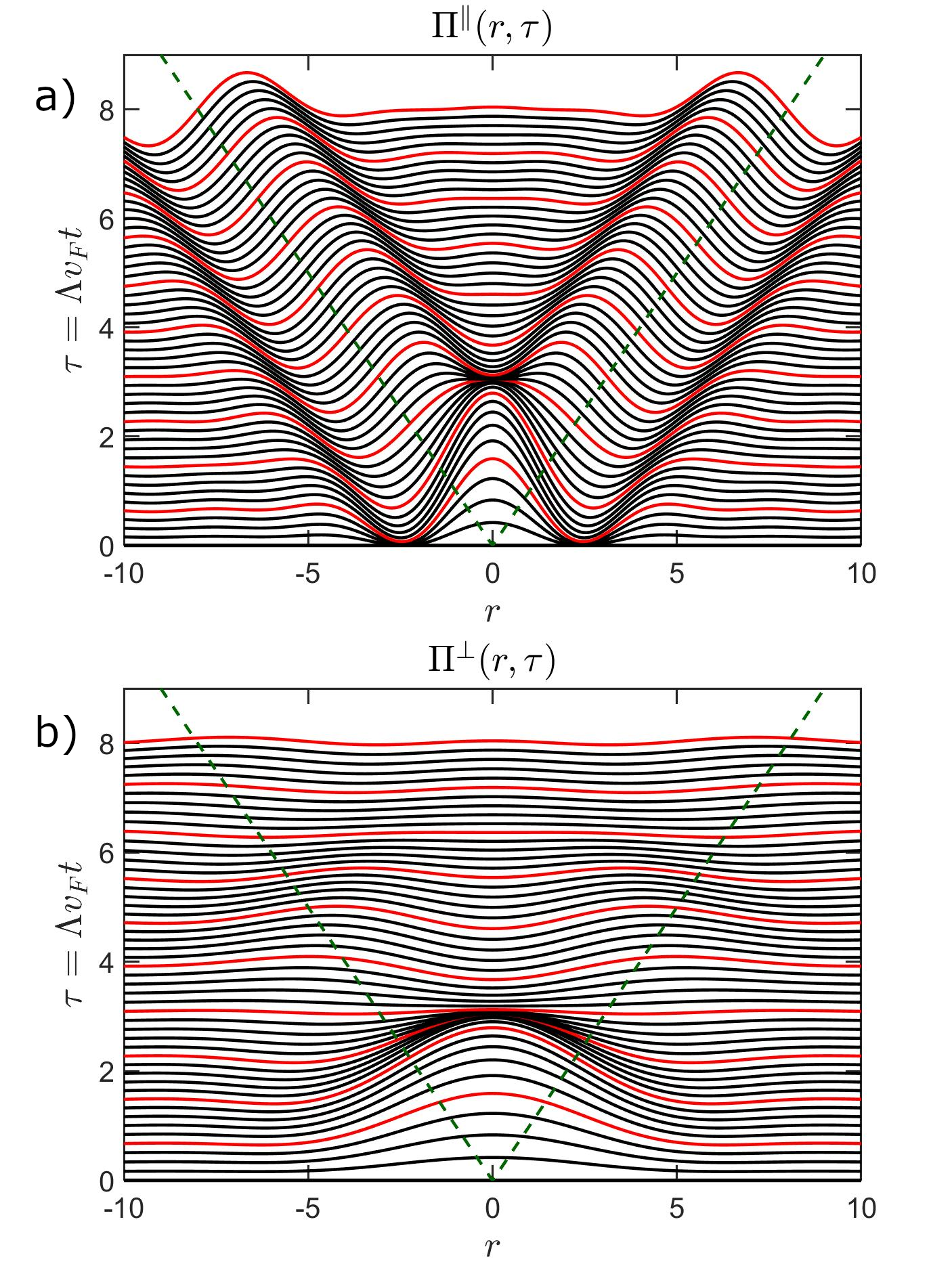}
		\caption{\label{fig:kuboxy} The spatial and temporal dependence of the  commutator is plotted in parallel ($a$) and perpendicular ($b$) case. The green dashed line represents the "light cone" with $v_\text{F}$ group velocity. The traces at fixed $\tau$ are shifted in the $y$ direction thus
				offering three-dimensional-like visualization. We set the chemical potential $\nu=0.1$.}
	\end{figure}

	In one dimension, we calculate the spatial dependence of the commutators analytically. Unlike the $\textbf{r}=0$ case, the shape of the correlation function depends on the spin component. Therefore we denote the parallel direction with $\Pi^{\parallel}(r,t)$ (when $\alpha=0,x$ in Eq.  \eqref{eq:kuboexp}) and $\Pi^{\perp}(r,t)$ in the perpendicular direction ($\alpha=y,z$) where $r$ denotes the spatial coordinate. The parallel and perpendicular correction functions expressed by the Green's functions are given by  
	
	\begin{eqnarray}
		\Pi^{\parallel}(r,t)=-2\text{Im}\big[G_{1}^{>}(t-r/v_\text{F})\bar{G}_{1}^{<}(t-r/v_\text{F})+\nonumber\\+(r \to -r)\big]
		\label{eq:resppara}\\
		\Pi^{\perp}(r,t)=-2\text{Im}\big[G_{1}^{>}(t-r/v_\text{F})\bar{G}_{1}^{<}(t+r/v_\text{F})+\nonumber\\+(r \to -r)\big]
		\label{eq:respperp}
		\end{eqnarray}
		The result is plotted on Fig. \ref{fig:kuboxy} for parallel and perpendicular case respectively. The short and late time behavior is obtained as  
	
	\begin{widetext}    
			\begin{eqnarray}
			\Pi^{\parallel}(r,t\to 0)\sim \frac{t}{r^3}\bigg[\Lambda r\cos\left(2\Lambda r\right)-\sin\left(2\Lambda r\right)-\Lambda r\cos\left(\Lambda r\right)\cos\left(\frac{\mu r}{v_\text{F}}\right)+\nonumber\\
			+\frac{\mu r}{v_\text{F}}\sin\left(\Lambda r\right)\sin\left(\frac{\mu r}{v_\text{F}}\right)+2\sin\left(\Lambda r\right)\cos\left(\frac{\mu r}{v_\text{F}}\right)\bigg]
			\label{eq:respparshort}\\
			\Pi^{\perp}(r,t\to 0)\sim \frac{t}{r^3}\left[
			\frac{\mu r}{v_\text{F}}\sin\left(\Lambda r\right)\sin\left(\frac{\mu r}{v_\text{F}}\right)+
			\Lambda r\cos\left(\Lambda r\right)\cos\left(\frac{\mu r}{v_\text{F}}\right)-\Lambda r\right]
			\label{eq:resppertshort}\\        
			\Pi^{\parallel/\perp}(r,t\to \infty)\sim \frac{\sin\left(\Lambda\left(v_\text{F}t-r\right)\right)}{t^2}\left[\cos\left(\mu\left(t\mp\frac{r}{v_\text{F}}\right)\right)-\cos\left(\Lambda\left(v_\text{F}t\mp r\right)\right)\right]+(r\to-r)
			\label{eq:respparlate}
			\end{eqnarray}
		\end{widetext}
	
		We find that the short time growth is linear in time and late time decay is $t^{-2}$ similarly that we obtained for $r=0$ case, but for late times the frequency of the oscillations is changed. This indicates that the temporal decay of the correlation function is independent from the spatial coordinate, and we expect this to hold in higher dimensions as well. 
	
	\section{\label{sec:otoc} Out-of-time-ordered commutator}
	
	The Dirac--Weyl systems are simple integrable models, thus  OTOC is expected to display a $\sim t^2$ initial growth and a power-law decay for late times. Using Eq. (\ref{eq:otocgen}), the OTOC for spatially non-separated spin operators is given by
	
	\begin{eqnarray}
	C_{\alpha\alpha}(t)= -\frac{1}{V_{d}^4}\sum_{\substack{k_1,k_2,\\ k_3}} \sum_{\substack{l_1,l_2,\\l_3}}T(k_1,k_2)T(l_1,l_2) \times \nonumber\\
	\times [\left \langle k_2|\sigma_\alpha|k_1\right\rangle
	\left \langle k_1|\sigma_\alpha|k_3\right\rangle
	\left \langle l_2|\sigma_\alpha|l_1\right\rangle
	\left \langle l_1|\sigma_\alpha|l_3\right\rangle 
	\langle a^{\dagger}_{k_2}a_{k_3}a^{\dagger}_{l_2}a_{l_3}\rangle\nonumber\\
	-\left \langle k_2|\sigma_\alpha|k_1\right\rangle
	\left \langle k_1|\sigma_\alpha|k_3\right\rangle
	\left \langle l_2|\sigma_\alpha|l_1\right\rangle
	\left \langle l_3|\sigma_\alpha|l_2\right\rangle
	\langle a^{\dagger}_{k_2}a_{k_3}a^{\dagger}_{l_3}a_{l_1}\rangle\nonumber\\
	-\left \langle k_2|\sigma_\alpha|k_1\right\rangle
	\left \langle k_3|\sigma_\alpha|k_2\right\rangle
	\left \langle l_2|\sigma_\alpha|l_1\right\rangle
	\left \langle l_1|\sigma_\alpha|l_3\right\rangle
	\langle a^{\dagger}_{k_3}a_{k_1}a^{\dagger}_{l_2}a_{l_3}\rangle\nonumber\\
	+\left \langle k_2|\sigma_\alpha|k_1\right\rangle
	\left \langle k_3|\sigma_\alpha|k_2\right\rangle
	\left \langle l_2|\sigma_\alpha|l_1\right\rangle
	\left \langle l_3|\sigma_\alpha|l_2\right\rangle
	\langle a^{\dagger}_{k_3}a_{k_1}a^{\dagger}_{l_3}a_{l_1}\rangle].\nonumber\\
	\label{eq:otoc}
	\end{eqnarray}    
	By using Wick's theorem, the expectation value of four fermionic operators is
	
	\begin{eqnarray}
	\left\langle a^{\dagger}_{k_1}a_{k_2}a^{\dagger}_{k_3}a_{k_4}\right\rangle=\delta_{k_1,k_2}\delta_{k_3,k_4}f(k_1)f(k_3)\nonumber\\
	+\delta_{k_1,k_4}\delta_{k_3,k_2}f(k_1)[1-f(k_3)].
	\label{eq:fourexp}
	\end{eqnarray}
	The commutator is split into two parts, the first and second one containing  $f(k_i)f(l_j)$ and $f(k_i)[1-f(l_j)]$ terms, respectively. The first part gives the square of the expectation value of the commutator, defined in Eq. (\ref{eq:kuboexp}), thus the OTOC looks as    
	
	\begin{equation}
	C_{\alpha\alpha}(t)= -\left\langle\left[\sigma_\alpha (t),\sigma_\alpha \right]\right\rangle^2+K_{\alpha\alpha}(t),    
	\label{eq:otocgen1}
	\end{equation}
	where $K_{\alpha\alpha}(t)$ is the variance of the commutator. It is given by    
	
	\begin{eqnarray}
	K_{\alpha\alpha}(t)=-\frac{1}{V^4}\sum_{\substack{k_1,k_2,\\ l_1,l_2}}
	\left \langle k_2|\sigma_\alpha|k_1\right\rangle
	\left \langle k_1|\sigma_\alpha|l_2\right\rangle\times\nonumber\\
	\times\left \langle l_2|\sigma_\alpha|l_1\right\rangle
	\left \langle l_1|\sigma_\alpha|k_2\right\rangle\times\nonumber\\
	\times \Big(T(k_1,k_2)T(l_1,l_2)[f(k_2)[1-f(l_2)]+f(k_1)[1-f(l_1)]]\nonumber\\
	-T(k_1,l_1)[f(k_2)(1-f(l_2))+f(l_2)(1-f(k_2))]\Big).\nonumber\\
	\label{eq:otoccorr}
	\end{eqnarray}    
	The product of the matrix elements is treated separately from remaining terms, and the integral over the angular variables is $N^{\alpha}_d=\left(M^{\alpha}_d\right)^2/2$ where $M^{\alpha}_d$ is defined in Eq. (\ref{eq:mat}). The remaining terms depend only the radial component of the momentum. Using Eq. (\ref{eq:greenless}) and Eq. (\ref{eq:greengreat}), we can rewrite the variance in Eq. (\ref{eq:otoccorr}) as
	
	\begin{eqnarray}
	K_{\alpha\alpha}(t)=2N^{\alpha}_d\Big(\left|G^{>}_{d}(t)+G^{<}_{d}(t)\right|^2\Big(\text{Re}\left[G^{>}_{d}(0)\bar{G}^{<}_{d}(0)\right]\nonumber\\
	-\text{Re}\left[G^{>}_{d}(t)\bar{G}^{<}_{d}(t)\right]\Big)-2\left(\text{Im}\left[G^{>}_{d}(t)\bar{G}^{<}_{d}(t)\right]\right)^2 \Big).\nonumber\\
	\label{eq:otoccorr2}
	\end{eqnarray} 
	Here $G^{</>}_{d}(0)=\lim_{t\to 0}G^{</>}_{d}(t)$. The detailed derivation is presented in Appendix \ref{app:appa}. Evaluating the integrals yields the variance as
	
		\begin{widetext}    
			\begin{gather}
				K_{\alpha\alpha}(\tau)=
				\begin{cases}
					\dfrac{32\Lambda^4}{(2\pi)^4}\left[\left(\dfrac{\sin(\tau)}{\tau}\right)^2\left[\dfrac{1-\nu^2}{2}-\dfrac{\cos(\tau)}{\tau}\left[\cos(\nu\tau)-\cos(\tau)\right]\right]-\left[\dfrac{\sin{\tau}}{\tau}\left[\cos(\nu\tau)-\cos(\tau)\right]\right]^2\right]        
					&\text{for } d=1,\vspace{2mm}\\
					\dfrac{\Lambda^8}{(2\pi)^4}\bigg[\left(\dfrac{\cos(\tau)+\tau\sin(\tau)-1}{\tau^2}\right)^2
					\Big[\dfrac{1-\nu^4}{4}-\dfrac{1}{\tau^4}\big[\left(\cos(\tau)+\tau\sin(\tau)\right)^2-(1+\nu^2\tau^2)-\\-\left(\tau\cos(\tau)-\sin(\tau)\right)^2-2(\cos(\tau)+\tau\sin(\tau))+2(1+\nu\tau(\tau\cos(\tau)-\sin(\tau)))\cos(\nu\tau)+\\
					+2(\nu\tau-(\tau\cos(\tau)-\sin(\tau)))\sin(\tau)\big]\Big]-\\
					-\dfrac{2}{\tau^8}\big[\left(\cos(\tau)+\tau \sin(\tau)-1\right)\left(\sin(\tau)-\tau\cos(\tau)-\left(\sin(\nu\tau)-\nu\tau\cos(\nu \tau)\right)\right)\big]^2\bigg]
					&\text{for } d=2,\vspace{2mm}\\
					\dfrac{16\Lambda^{12}}{(2\pi)^8}\bigg[\left(\dfrac{(\tau^2-2)\sin(\tau)+2\tau\cos(\tau)}{\tau^3}\right)^2\Big[\dfrac{1-\nu^6}{9}-\dfrac{1}{\tau^6}\big[4\tau(\tau^2-2)\sin(2\tau)-\\
					-(\tau^4-8\tau^2+4)\cos(2\tau)-(4+\nu^4\tau^4)+2((\tau^2-2)\cos(\tau)-2\tau\sin(\tau))\times\\
					\times((\tau^2\nu^2-2)\cos(\nu\tau)-2\nu\tau\sin(\nu\tau))
					\big]\Big]-\dfrac{2}{\tau^{12}}\Big[\left(2\tau\cos(\tau)-(2-\tau^2)\sin(\tau)
					\right)\times\\
					\times\Big((2-\tau^2)\cos(\tau)+2\tau\sin(\tau)-\left((2-\nu^2\tau^2)\cos(\nu\tau)+2\tau\sin(\nu\tau)\right)    \Big) \Big]^2\bigg] &\text{for } d=3.    
				\end{cases}
				\label{eq:otoce}    
			\end{gather}        
	\end{widetext}
	Here, we used again the dimensionless variables for time ($\tau$) and the chemical potential ($\nu$).
	
\begin{figure}[!ht]
		\includegraphics[width=\linewidth]{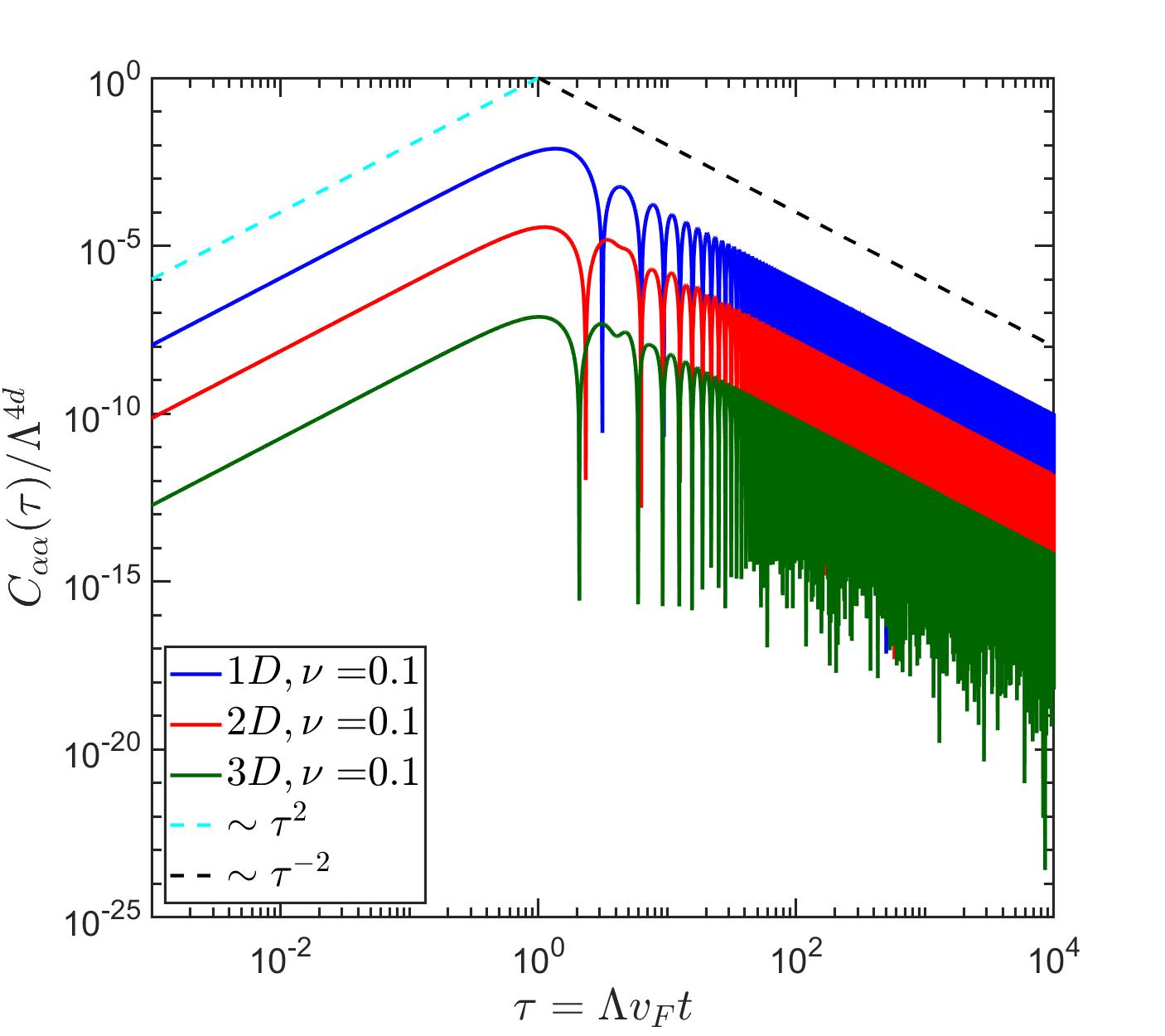}
		\caption{\label{fig:otocshort}  Short and late time behavior of the OTOC in one (blue), two (red) and three (green) dimensions at $\nu=0.1$. For short times, OTOC exhibits $\tau^2$ growth with a chemical potential dependent prefactor. For intermediate and late times it decays as $\tau^{-2}$ power-law.}
	\end{figure}

	Eq. \eqref{eq:otoce} allows us to investigate the short and late time behavior of the OTOC, similarly to the case of the commutator in Eq. \eqref{eq:kuboe}. The natural energy scales in the problem are $\mu$ and $\Lambda v_\text{F}$, translating into three separate temporal windows for short, intermediate and late times as $\tau\ll 1$, $1\ll\tau\ll 1/\nu$ and $1/\nu\ll\tau$, respectively, but similarly to the simple commutator, only the cutoff related timescale matters. For short times, the OTOC grows with $t^2$ also shown in Fig. \ref{fig:otocshort}. This behavior follows from a Baker-Campbell-Hausdorff expansion of $\sigma_\alpha(t)$ with the nested commutators \cite{Dra2017,Roberts2016}
	
	\begin{equation}
	\sigma_\alpha(t)=\sigma_\alpha+it \left[H,\sigma_\alpha\right]+\frac{(it)^2}{2!}\left[ H,\left[H,\sigma_\alpha\right]\right]+\dots
	\label{eq:hausdorf}
	\end{equation}
	The contribution of the first term in Eq. (\ref{eq:hausdorf}) gives trivially vanishing contribution to both the expectation value of the commutator and the OTOC. The $t^2$ growth arises from the second term in Eq. (\ref{eq:hausdorf}) with the coefficient $\langle\left[\left[H,\sigma_\alpha\right],\sigma_\alpha\right]^2\rangle$. This short time growth originates from both the  square of the expectation value of the commutator and the variance of the commutator in Eq. (\ref{eq:otoce}) and it is given by
	
	\begin{eqnarray}
		C_{\alpha\alpha}(t\to 0)\sim \Lambda^{2d}t^2((\Lambda v_\text{F})^{d+1}-\mu^{d+1})\times\nonumber\\
		\times((\Lambda v_\text{F})^{d+1}-A_d\mu^{d+1})
		\label{eq:otocshort},
		\end{eqnarray}
		where $A_1=5/13$, $A_2=5/11$ and $A_3=29/61$, which comes from the fact that both the square of the commutator and the variance give finite contribution to the initial growth. Similarly to the short time behavior of the correlation function in \eqref{eq:kuboshort} the exponent of the cutoff and chemical potential come from the leading order expansion of the time evolution operator. 
	We note that the second term in Eq. (\ref{eq:hausdorf}) is also responsible for the linear growth of the response function for short times and it ensures that both the linear response and the OTOC have to be real since $H$ and $\sigma_\alpha$ are Hermitian operators, thus the expectation value of their commutator is also real. 
	
	For intermediate and late times, the OTOC decays in an identical power-law fashion, shown in Fig. \ref{fig:otocshort}. Irrespective of the value of $\mu$, the asymptotics of the temporal decay looks as
	
	\begin{equation}
		C_{\alpha\alpha}(t\to \infty)\sim\frac{\Lambda^{2(d-1)}\sin^2(\Lambda v_\text{F}t)}{t^2}\left((\Lambda v_\text{F})^{2d}-\mu^{2d}\right)
		\label{eq:otoclate}
		\end{equation}
	It decays as $t^{-2}$, similarly to the simple commutator and is independent of the chemical potential and the spatial dimension. The coefficient is determined by the difference between the square of the number of states at the cutoff energy and the chemical potential. This behavior originates from the time-independent part of Eq. (\ref{eq:otoccorr2}) i.e., $\text{Re}\left[G^{>}_{d}(0^+)\bar{G}^{<}_{d}(0^+)\right]$, the remaining terms of the OTOC only give subleading $t^{-4}$ decay.  Our results agree with those in Ref.[\onlinecite{Lin20182}].
	
 We again argue that the $t^{-2}$ late time decay of the OTOC represent the typical response of Dirac--Weyl fermions. The OTOC is decomposed as the sum of two
terms in Eq. \eqref{eq:otocgen1}, namely the square of the expectation value of the commutator and its variance. The former gives only a subleading $t^{-4}$ decay from
Eq. \eqref{eq:kubolate}. The latter, however, gives the dominant $t^{-2}$ decay for late times. In order to see this, we again focus on the characteristic linear 
energy-momentum relationship of Dirac-Weyl fermions, i.e. $\varepsilon_{\lambda}(\mathbf{k})\sim |\mathbf{k}|$.
By keeping the momentum dependence of the time dependent factors $T(k,l)$ in the variance in Eq. \eqref{eq:otoccorr} and neglecting the wavevector dependence of the other terms,
we obtain the aforementioned $t^{-2}$ decay after the $d$-dimensional momentum integrals. This is therefore the characteristic feature of the OTOC in Dirac--Weyl systems.

\begin{figure}[!ht]
		\includegraphics[width=\linewidth]{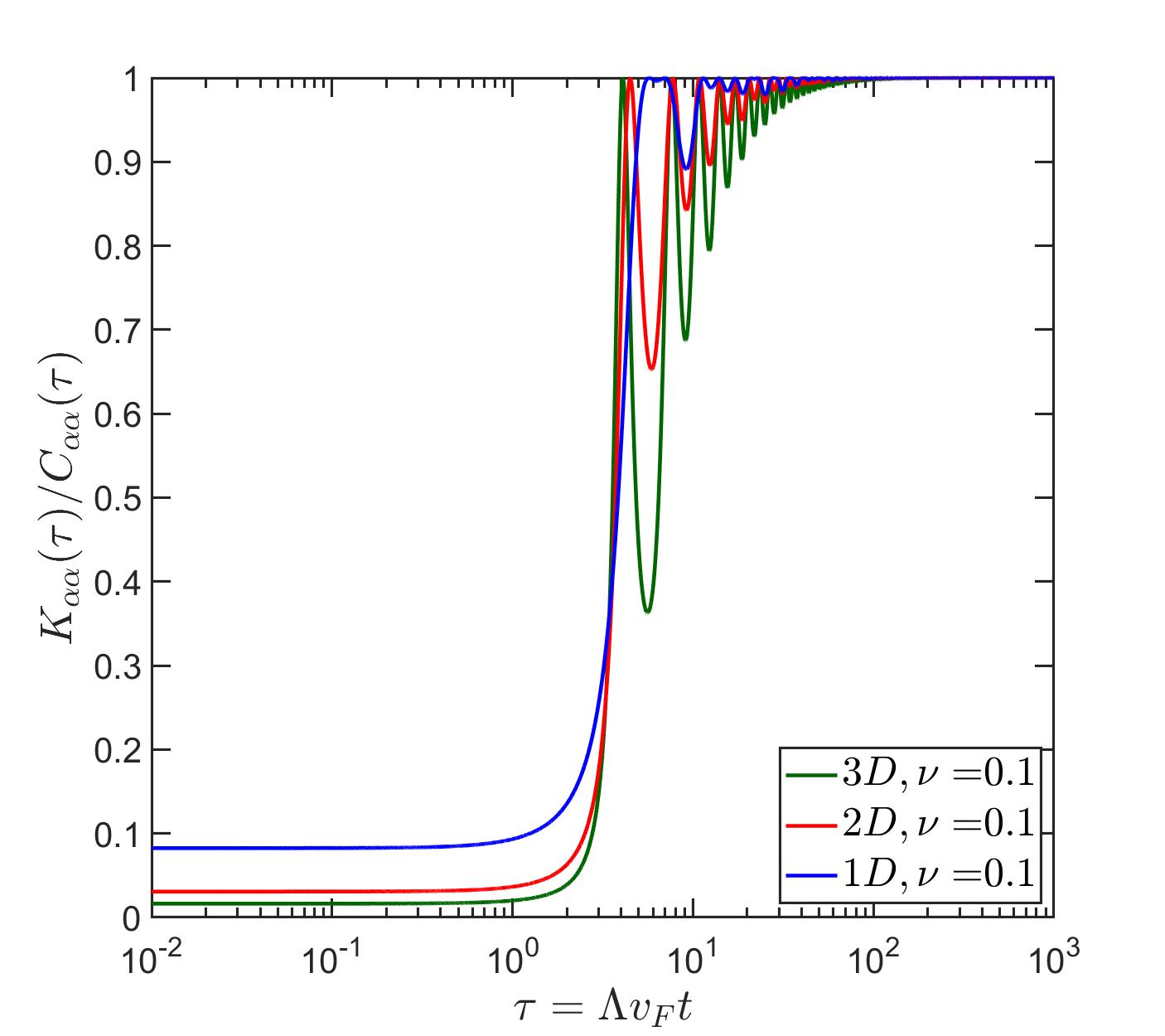}
		\caption{\label{fig:var} Ratio of the variance and OTOC as a function of time at $\nu=0.1$. For $\tau \ll 1 $, the main contribution to OTOC comes from the square of the expectation value of the
				commutator, while when $\tau \gg 1 $ the variance dominates.}
	\end{figure}

	The influence of the variance on  OTOC is plotted in Fig. \ref{fig:var}. For short times, both the square of the expectation value of the commutator and the variance scales with $t^2$ but former parametrically is dominant over the latter. This is seen in Fig. \ref{fig:var} since the $K_{\alpha\alpha}(\tau)/C_{\alpha\alpha}(\tau)$ ratio is small for $\tau \ll 1$. 
	For late times, the ratio tends to one, independently of the chemical potential, indicating that the OTOC is dominated by the variance in this regime.
	
\begin{figure}[!ht]
		\includegraphics[width=\linewidth]{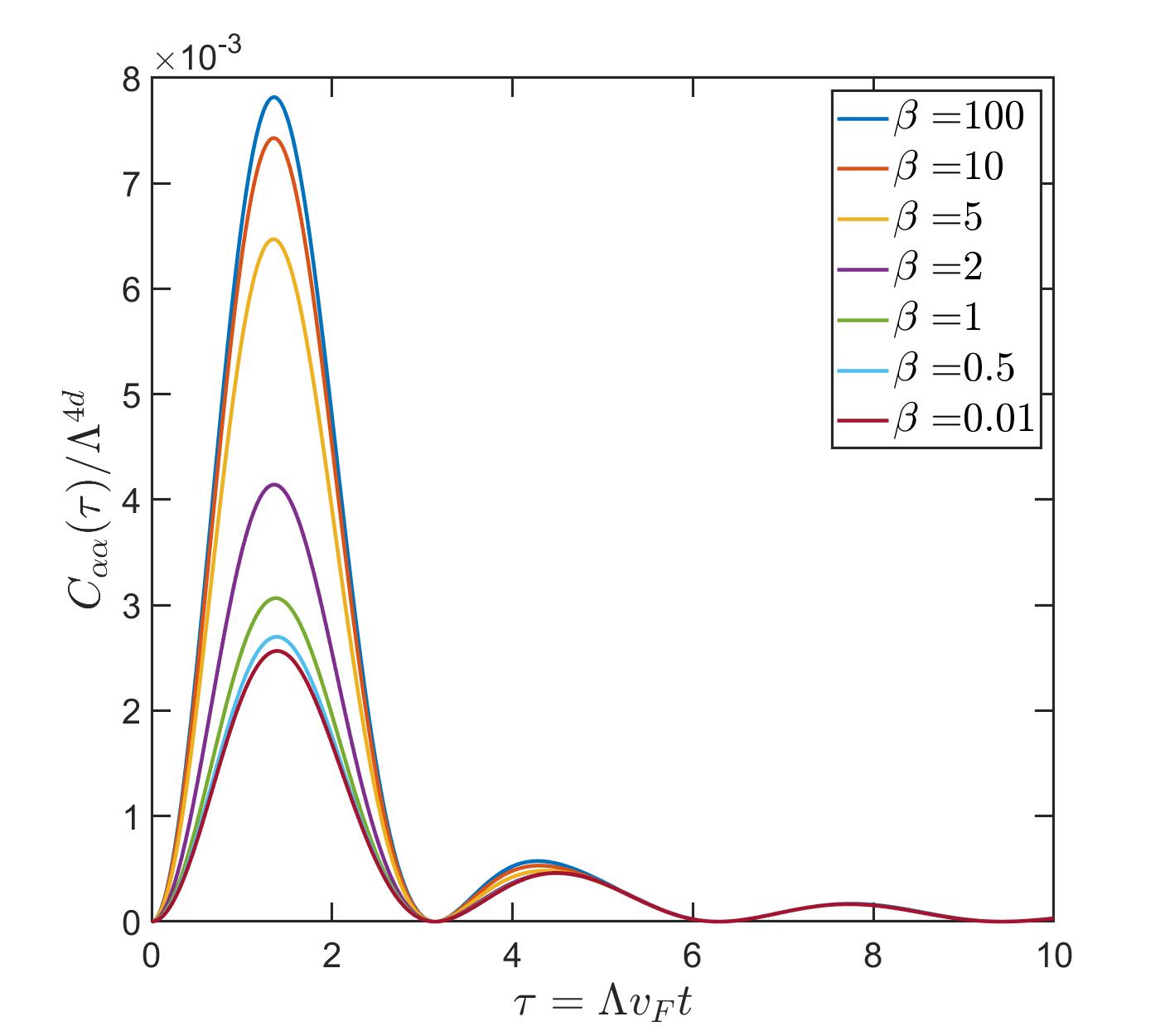}
		\caption{\label{fig:otoctemp} The temporal dynamics of the spin and density OTOC is plotted for different temperatures in one  dimension for $\nu=0$. $\beta=\Lambda v_\text{F}/k_{B}T$ denotes the inverse temperature. The main features of $C_{\alpha\alpha}(\tau)$ remain the same with increasing temperature, but the amplitude decreases. At the $\beta=0$ limit the OTOC remains finite. }
	\end{figure}

	We also investigated the temperature dependence of the OTOC similarly to the simple commutator, but unlike the simple commutator, the OTOC remains finite at infinite temperatures. This limit is expressed with the Green's functions as
	
	\begin{equation}
		C_{\alpha\alpha}(t)= 4\left(M^\alpha_d\right)^2G^2_d(t)\left[G^2_d(0)-G^2_d(t)\right],
		\label{eq:otoctemp}
		\end{equation}
		where $G_d(t)=\lim_{T\to \infty}G_{d}^{<}(t)=\lim_{T\to \infty}G_{d}^{>}(t)$, i.e. the two Green's functions are identical at infinite temperature. In this case, only the variance gives non-zero contribution to the OTOC but the correlation function vanishes. The temperature dependence of the OTOC, obtained from evaluating the integrals in Green's function numerically, is displayed in Fig. \ref{fig:otoctemp}. The initial growth and late time decay retain the original $t^2$ and $t^{-2}$ behavior, only the amplitude decreases smoothly with increasing temperature. Note that the late time behavior of OTOC is independent of temperature in contrast to the simple  commutator. Calculating the asymptotic time dependence we get $C_{\alpha\alpha}(t) \sim  \sin^2(\Lambda v_\text{F}t)/t^2$ which agrees with Eq. \eqref{eq:otoclate} for $T=0$. 
	
\begin{figure}[!ht]
		\includegraphics[width=\linewidth]{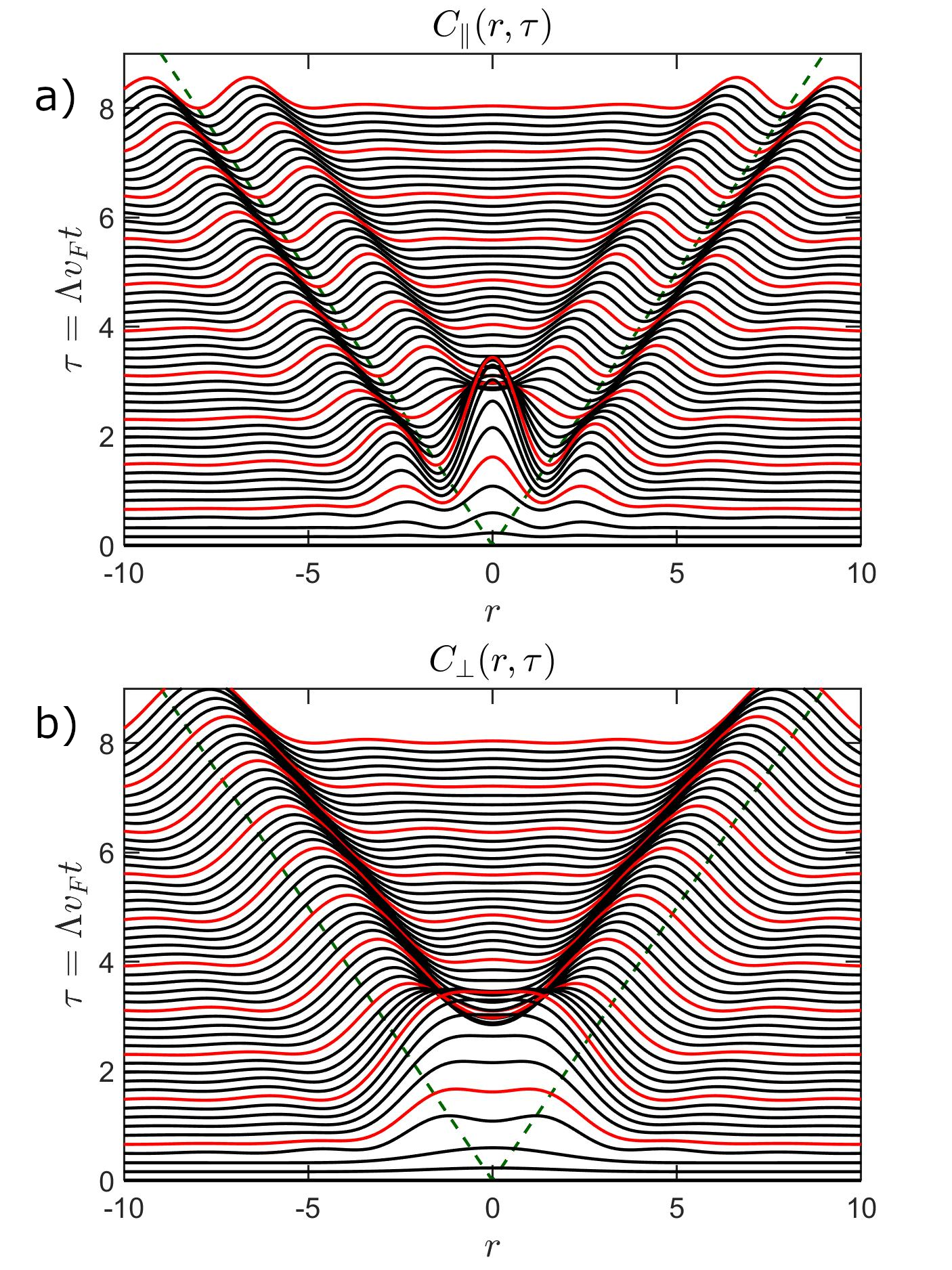}
		\caption{\label{fig:otocxy} The spatial and time dependent OTOC is plotted in parallel ($a$) and perpendicular ($b$) case. The green dashed line represents the "light cone" with $v_\text{F}$ group velocity. The traces at fixed $\tau$ are shifted in the $y$ direction thus     offering three-dimensional-like visualization. The decay inside the light cone is a characteristic feature of slow scramblers. We set the chemical potential $\nu=0.1$.}
	\end{figure}

	In one dimension, we compute the spatial dependence of the OTOC at zero temperature, which is taken into account by inserting  $\exp{(i(k_1-k_2+l_1-l_2)r)}$ to Eq. \eqref{eq:otoc}. Similarly to the simple commutator, due to the product of the matrix elements, we distinguish parallel and perpendicular cases. The obtained results are plotted in Fig. \ref{fig:otocxy}. For short times, it displays $t^2$ growth while for late times, inside the light cone, the parallel and the perpendicular directions decay identically as
	
	\begin{multline}
			C_{\parallel/\perp}(r,t\to \infty)\sim\frac{\sin^2(\Lambda(v_\text{F}t-r))}{t^2}\left((\Lambda v_\text{F})^{2}-\mu^{2}\right)\\+\left(r\to -r\right).
			\label{eq:otoclatex}
		\end{multline}
		Based on our results, we can assume that the OTOC behaves similarly in higher dimensions if we take into account the spatial dependence.
	
	In general, the quantum butterfly effect shows up in systems with exponential growth of the OTOC at short times and a large late time value\cite{Roberts20152,Roberts2016,Maldacena2016}. For non-interacting Dirac-Weyl systems, the OTOC scales with $t^2$ initially, i.e. with the lowest possible power and no exponential growth is identified.
	The late time OTOC vanishes which agrees with the general expectations since our models are non-interacting without any chaotic feature. This means the information encoded in local operators is lost slowly through time evolution\cite{Huang2016,Chen2016}. The same short time behavior was identified in different integrable models like Heisenberg XXZ chain\cite{Dra2017}, quantum Ising chain\cite{Lin2018}, and XY chain\cite{Jiahui2019}. These models also exhibit late time power-law decay, similarly to our findings, though the exponents depend on the actual system.     
	
	\section{\label{sec:summary}Summary}
	
	In this work, we have investigated the dynamics of the expectation value of the spin and density commutators and the corresponding OTOC in Dirac--Weyl systems in $d=1$, $2$ and $3$ dimensions. These operators contain useful information about the electric, magnetic and transport properties since the electric current operator is proportional to the spin. We use a sharp cutoff scheme throughout which is rather common in condensed matter physics in e.g. tight-binding models, such as for graphene. The problem at hand features two energy scales the cutoff and the chemical potential, thus we expect three different temporal regions, namely short, intermediate and late times. However, we found that only the cutoff related timescale matters and  only two distinct temporal regions need to be considered.

	For the correlation function, we obtained a linear initial rise in time. In the late time regime, the commutator decays with  $t^{-2}$ and its prefactor is proportional to the difference between the density of states at the cutoff energy and the chemical potential. We found that short and late time behavior does not change with the spatial dependence in one dimension, and we expect the same for higher dimensions. 
For finite temperatures the expectation value of the commutator has the same temporal behavior, although the magnitude is decreasing and at the $T\to \infty$ limit, it completely vanishes.
	
	The OTOC of local operators displays robust behavior, mostly independent of chemical potential. For short times, the OTOC grows with $t^2$, 
the lowest possible power, while for late times, it decays with $t^{-2}$ for all spin components, but its prefactor is determined by the square of 
the number of states instead of the density of the states. The short time behavior is determined by both the square of the expectation value of the 
commutator and the variance while for late times the variance gives the dominant part. The $t^{-2}$ decay for late times is identified as the characteristic signature
of Dirac-Weyl systems and follows from the linear energy-momentum relation.
The OTOC remains finite at infinite temperature since the variance of the commutator contains terms that are independent of temperature. The short and late time behaviors are not altered by the spatial coordinate.
	These results altogether indicate that these systems are slow information scramblers. Our findings are essential when the effect of other sources of information scrambling (interaction, disorder) are analyzed on top of the non-interacting results.
	
	\begin{acknowledgments}
		This research is supported by the National Research, Development and Innovation Office - NKFIH within the Quantum Technology National Excellence Program (Project No. 2017-1.2.1-NKP-2017-00001) and K119442, SNN118028 and by the BME-Nanotechnology FIKP grant of EMMI (BME FIKP-NAT) and by the \'{U}NKP-19-3 New National Excellence Program of the Ministry for Innovation and Technology
	\end{acknowledgments}
	
	\appendix
	\section{\label{app:appa} Derivation of variance for the OTOC}
	
	To obtain the expectation value of the spin commutator and the OTOC, first, we evaluate the integrals defined by  $G_{d}^{</>}(t)$ in Eqs. (\ref{eq:greenless}) and (\ref{eq:greengreat}). We focus on the zero temperature case in one dimension but higher-dimensional extensions are straightforward. It looks as     
	
		\begin{eqnarray}
		G_{1}^{<}(t)=\sum_{\lambda}\int_{0}^{\Lambda}\dfrac{dk}{2\pi}~e^{-i\lambda v_\text{F}kt}\Theta(\mu-\varepsilon_{\lambda}(k)),
		\label{eq:greenless2}\\
		G_{1}^{>}(t)=\sum_{\lambda}\int_{0}^{\Lambda}\frac{dk}{2\pi}~e^{-i\lambda v_\text{F}kt}\Theta(\varepsilon_{\lambda}(k)-\mu).
		\label{eq:greengreat2}
		\end{eqnarray}
	Assuming $\mu>0$, the explicit form of Green's functions with the dimensionless variables are
	
		\begin{eqnarray}
		G_{1}^{<}(\tau)=\frac{\Lambda}{2\pi}\frac{i}{\tau}\left[e^{-i\nu\tau}-e^{i\tau}\right],
		\label{eq:greenless3}\\
		G_{1}^{>}(\tau)=\frac{\Lambda}{2\pi}\frac{i}{\tau}\left[e^{-i\tau}-e^{-i\nu\tau}\right].
		\label{eq:greengreat3}
		\end{eqnarray}
	The variance term of OTOC  can be expressed with $G_{d}^{</>}(t)$ similarly to the correlation function in Eq. (\ref{eq:kubogreen}). Rewriting the wave vector integral with spherical coordinates, we can separate the angular integrals since the matrix elements are not containing the radial part of the integral variables. Integrating over the matrix elements gives $N^{\alpha}_d$  which is independent from the band indices. The remaining parts are the Fermi function and the terms containing time evolution which are depending on the band index and the radial component of the momentum  vector through the energy. These terms of Eq. (\ref{eq:otoccorr}) can be split into two parts as     
	\begin{eqnarray}
	I_1=T(k_1,k_2)T(l_1,l_2)\big(f(k_2)[1-f(l_2)]+\nonumber\\
	+f(k_1)[1-f(l_1)]\big)
	\label{eq:term1}
	\end{eqnarray}
	
	\begin{equation}
	I_2=T(k_1,l_1)[f(k_2)(1-f(l_2))+f(l_2)(1-f(k_2))]
	\label{eq:term2}
	\end{equation}
	Since $I_1$ contains the product of two time dependent terms and products of Fermi functions, we rewrite it as	
	\begin{eqnarray}
	I_1=-T(k_1,k_2)T(l_1,l_2)[f(k_2)-f(k_1)][f(l_2)-f(l_1)]+\nonumber\\
	+T(k_1,k_2)T(l_1,l_2)\left[f(k_2)(1-f(l_1))+f(k_1)(1-f(l_2))\right]\nonumber\\
	\label{eq:term12}
	\end{eqnarray} 
	If we integrate over the radial component of the momentum and make the summations over the band indices the result matches with the square of radial integrals in the commutator in Eq. (\ref{eq:kuboexp}). The integral of second time evolution containing term in Eq. (\ref{eq:term12}) is also traced back to the product of Green's functions. Thus the summation over the band indices and the integrals of $I_1$ is written with the Green's functions as	
	\begin{eqnarray}
	\sum_{\substack{\lambda_1,\lambda_2,\\ \mu_1,\mu_2}}\int_{0}^{\infty}\dfrac{dk_1}{2\pi}\int_{0}^{\infty}\dfrac{dk_2}{2\pi}\int_{0}^{\infty}\dfrac{dl_1}{2\pi}\int_{0}^{\infty}\dfrac{dl_2}{2\pi}\times \nonumber\\
	\times k^{d-1}_1 k^{d-1}_2 l^{d-1}_1 l^{d-1}_2 I_1 =\nonumber\\
	-\left(2i\text{Im}\left[G_{d}^{>}(t)\bar{G}_{d}^{<}(t)\right]\right)^2
	+2\text{Re}\left[G^{>}_{d}(t)\bar{G}^{<}_{d}(t)\right]\times \nonumber\\
	\times\sum_{\lambda_1,\mu_2}\int_{0}^{\infty}\dfrac{dk_1}{2\pi}\int_{0}^{\infty}\dfrac{dl_2}{2\pi}k^{d-1}_1l^{d-1}_2~ T(k_1,l_2).\nonumber\\
	\label{eq:term13}
	\end{eqnarray} 
	We swapped variables in the last term of Eq. (\ref{eq:term12}). The remaining two integrals gives the same result as $\left|G^{>}_{d}(t)+G^{<}_{d}(t)\right|^2$. 
	
	The formula of $I_2$ in Eq. (\ref{eq:term2}) resembles closely to the second term of Eq. (\ref{eq:term12}) except it has only one time dependent part. Thus, if we multiply it with $\lim_{t\to 0^{+}}T(k_2,l_2)$ which is $1$  we get  the same structure as in Eq. (\ref{eq:term12}). This implies that the result is the same as the last term of Eq. (\ref{eq:term13}) after the integrals and summations. Taking the $t\to 0$ limit, we end up with the last term of Eq. (\ref{eq:otoccorr}) expressed with Green's functions as
	
	\begin{eqnarray}
	\sum_{\substack{\lambda_1,\lambda_2,\\ \mu_1,\mu_2}}\int_{0}^{\infty}\dfrac{dk_1}{2\pi}\int_{0}^{\infty}\dfrac{dk_2}{2\pi}\int_{0}^{\infty}\dfrac{dl_1}{2\pi}\int_{0}^{\infty}\dfrac{dl_2}{2\pi}\times \nonumber\\
	\times k^{d-1}_1 k^{d-1}_2 l^{d-1}_1 l^{d-1}_2 I_2=\nonumber\\
	2\left|G^{>}_{d}(t)+G^{<}_{d}(t)\right|^2\text{Re}\left[G^{>}_{d}(0^+)\bar{G}^{<}_{d}(0^+)\right]
	\label{eq:term22}
	\end{eqnarray}
	Substituting the result of (\ref{eq:term13})  and (\ref{eq:term22}) into Eq. (\ref{eq:otoccorr}) and multiply it with $N^{\alpha}_d$ from the angular integrals, we obtain the variance with the Green's functions in Eq. (\ref{eq:otoccorr2}). By inserting the explicit formula of the corresponding Green's function, we obtain the complete time dependence of the OTOC.

	\bibliographystyle{apsrev}
	\bibliography{otoc}
	
\end{document}